\documentstyle[12pt,epsfig,amsmath,amssymb,pstricks]{article}
\begin{document}

\title{\bf Information contents and architectural requirements of observer ``ready'' states}

\author{{Chris Fields}\\ \\
{\it 21 Rue des Lavandi\`eres}\\
{\it Caunes Minervois 11160 France}\\ \\
{fieldsres@gmail.com}}
\maketitle

\begin{abstract}
A functional analysis of the task of observing multiple macroscopic quantum systems over an extended period of time and then reporting the accumulated results is used to investigate the information that must be encoded in the ``ready'' state $|\mathcal{O}^{\mathit{r}}\rangle$ of any finite, macroscopic observer $\mathcal{O}$ capable of performing this task.  Decoherence considerations show that this task can be considered as involving local observations under classical conditions (LOCC), allowing the use of classical automata theory to define a minimal observer.  It is shown that such a minimal observer must implement a functional architecture equivalent to a classical Turing machine and must encode in $|\mathcal{O}^{\mathit{r}}\rangle$ a classical specification of the complete set of reportable apparatus states.  The observation task is then re-characterized employing an explicit model of such a minimal observer, and it is shown that both the assumption that external systems have well-defined boundaries against the environment and the assumption of decoherence are unnecessary for the characterization of measurements made by a minimal observer.  It is shown that the observables available to a minimal observer are positive operator-valued measures (POVMs) and that the measurement results reported by a minimal observer comply with the Born rule.  The differences in underlying physical assumptions between this ``systems-free'' treatment of observation and that traditionally employed in analyses of quantum measurement and quantum communication are discussed.  
\end{abstract}

\textbf{Keywords:}  Measurement; minimal observer; Born rule; decoherence; Turing-equivalent architecture; quantum Darwinism.  PACSS: 03.65.Ca; 03.65.Ta; 03.65.Yz

\section{Introduction}

Consider a finite, macroscopic quantum system $\mathcal{O}$, an ``observer'', that interacts via a common channel $\mathcal{E}$, the shared environment, with $N$ macroscopic quantum systems $\mathcal{A}_{\mathit{i}}$.  Assume the systems $\mathcal{A}_{\mathit{i}}$ are sufficiently separated to be mutually decoupled, and that each serves as an ``apparatus'' in the sense that only a small projection of its overall state, which to avoid ambiguity will be referred to as its ``read-out'' component $\mathcal{R}_{\mathit{i}}$, is of interest to $\mathcal{O}$.  The task of $\mathcal{O}$ is to observe the $\mathcal{A}_{\mathit{i}}$ for an extended period of time, long enough for each of the $\mathcal{A}_{\mathit{i}}$ to undergo numerous state transitions, and then to report the states of the $\mathcal{R}_{\mathit{i}}$ that have been observed.  This paper addresses two questions about any physical system $\mathcal{O}$ capable of carrying out this task.  First, what information must $\mathcal{O}$ encode at the outset of the period of observation, i.e. in a ``ready'' state $|\mathcal{O}^{\mathit{r}}\rangle$, in order to complete this task successfully?  Second, into what functional architecture must this prior information be organized to enable $\mathcal{O}$ both to carry out the observations and to report the results?  It shows that these questions can be answered precisely, and that the answers have significant consequences for the physical description of measurement.  It shows, in particular, that quantum measurement can be completely described within a ``systems-free'' formulation that does not assume well-defined boundaries between the $\mathcal{A}_{\mathit{i}}$ and the surrounding environment $\mathcal{E}$ and does not depend on either the physical assumptions or the formalism of decoherence.

The question of what information observers must encode in order to conduct observations and report the results has generally been avoided in discussions of quantum measurement.  In a widely-cited review of decoherence, Zurek \cite{zurek03rev} remarks that observers differ from apparatus only in their ability to ``readily consult the content of their memory'' (p. 759), but nowhere specifies either what memory contents are consulted or what memory contents might be required.  In the formal context of the ``environment as witness'' \cite{zurek04, zurek05} and quantum Darwinism \cite{zurek06, zurek07grand, zurek09rev} programs, prior knowledge on the part of observers and even observers themselves appear superfluous; the pointer states of all possible systems of interest are einselected by decohering interactions and encoded by the environment at much higher redundancy and hence greater robustness against decay than could be achieved by any finite observer.  Observers in this context are merely localized recipients of information, their state transitions driven by the environment in the same way that they are driven by the apparatus in the traditional representation of the von Neumann chain (e.g. \cite{schloss04, schloss07}).  In reformulations of measurement as a quantum communication process, the state of Bob the receiver is similarly treated as driven by the quantum communication channel through which Alice the sender transmits messages (e.g. \cite{galindo01, griffiths07}); while Bob must sometimes perform operations to decode a message, the instructions for performing these operations also come, via a parallel classical channel, from Alice.  Such passive observers can be altogether replaced by sequences of conditional probabilities of potential measurement outcomes; doing so clearly obviates any questions regarding what information observer states might encode, and is often considered a clarification of quantum measurement (e.g. \cite{goldstein98}).  Even when quantum states are considered fully ``subjective'' and measurement is conceptualized as a process of belief revision by agents faced with new information, the question of what information an agent must encode to enable such belief revision is considered otiose: ``it is not the task of [the] theory to explain how the transition [it] signifies comes about within the mind of the agent'' (\cite{fuchs02}, p. 1015).

This paper shows that neglecting the requirements placed on the observer by the pragmatic \textit{task} of observation has both obscured the role of decoherence in the environmental encoding of pointer states and needlessly complicated the analysis of quantum measurement.  The remainder of the paper is organized as follows.  In Section 2, the physical assumptions that must be made in order to characterize the transfer of information from a macroscopic apparatus $\mathcal{A}$ to a macroscopic observer $\mathcal{O}$ are examined using the standard language of decoherence.  The roles played by the assumptions of einselection and that the environment $\mathcal{E}$ serves as a ``witness'' that encodes the pointer states of $\mathcal{A}$ are reviewed.  It is shown that under these assumptions, measurement interactions can be characterized as local observations with classical communication (LOCC), and hence as resulting in $\mathcal{O}$ possessing localized classical information.  A functional decomposition of the task of observing multiple macroscopic systems $\mathcal{A}_{\mathit{i}}$ under LOCC conditions is used to define both the minimal classical information that must be encoded by the ready state $|\mathcal{O}^{\mathit{r}}\rangle$ and the minimal functional architecture that must be implemented by any finite observer $\mathcal{O}$ capable of observing the read-outs $\mathcal{R}_{\mathit{i}}$ of any finite set of macroscopic quantum systems $\mathcal{A}_{\mathit{i}}$ embedded in an environment $\mathcal{E}$.  The ``minimal observer'' $\mathcal{O}$ defined by this task analysis has, not surprisingly, the capabilities and structure of a typical laboratory data-collection system implemented on a typical laboratory computer.  In Section 3, the physical assumptions made in characterizing the task of observing multiple systems $\mathcal{A}_{\mathit{i}}$ are progressively relaxed, and it is shown that the minimal observer $\mathcal{O}$ can be represented as interacting with the quantum environment $\mathcal{E}$ without the need for any assumptions regarding the external systems $\mathcal{A}_{\mathit{i}}$ or their interactions with $\mathcal{E}$.  The observables available to $\mathcal{O}$ in this ``systems-free'' formulation are defined and shown to be positive operator-valued measures (POVMs).  A variation of Zurek's proof from envariance \cite{zurek05env} is used to show that observations made with these observables obey the Born rule.  This proof employs only the systems-free formulation and is independent of any assumptions regarding decoherence, einselection, or the interaction of external systems with the environment; hence it shows that \textit{classical} observers can be expected to report results consistent with quantum mechanics solely in virtue of their functional architectures and the information encoded in $|\mathcal{O}^{\mathit{r}}\rangle$.  The implications of these facts for the physical interpretations of both quantum and classical mechanics are discussed in Sect. 4.

\section{Task analysis defines a minimal observer}

The fundamental insight of decoherence theory is that observers inhabit an environment, and that the transfer of information from a quantum system $\mathcal{A}$ to an observer $\mathcal{O}$ - that is, measurement - therefore requires that $\mathcal{A}$ be open to interaction with this environment \cite{zurek03rev, schloss04, schloss07}.  For the present purposes, both $\mathcal{A}$ and $\mathcal{O}$ are assumed to be macroscopic, and the environment $\mathcal{E}$ is assumed to be large enough to contain many such systems and observers and to have sufficiently many degrees of freedom to impose strong decoherence.

Interposing a strongly-decohering environment between apparatus and observer replaces the traditional von Neumann chain:

\begin{equation}
(\sum_{k}\lambda_{k}|s_{k}\rangle)|\mathcal{A}^{\mathit{i}}\rangle|\mathcal{O}^{\mathit{r}}\rangle\,\rightarrow \mathit{\sum_{k}\lambda_{k}|s_{k}\rangle|a_{k}\rangle|o_{k}\rangle\,\rightarrow |s_{f}\rangle|a_{f}\rangle|o_{f}\rangle}
\end{equation}

with its implications of direct system-apparatus-observer entanglement followed by non-unitary ``collapse'' or unitary ``branching'' into a final joint eigenstate $|s_{f}\rangle|a_{f}\rangle|o_{f}\rangle$ (e.g. \cite{schloss04, schloss07}) by two decoupled processes:

\begin{equation}
(\sum_{k}\lambda_{k}|s_{k}\rangle)|\mathcal{A}^{\mathit{i}}\rangle|\mathcal{E}^{\mathit{i}}\rangle\,\rightarrow \mathit{\sum_{k}\lambda_{k}|s_{k}\rangle|a_{k}\rangle|e_{k}\rangle\,\rightarrow |s_{f}\rangle|a_{f}\rangle|e_{f}\rangle}
\end{equation}

and

\begin{equation}
|e_{f}^{\prime}\rangle|\mathcal{O}^{\mathit{r}}\rangle\,\rightarrow \mathit{|e_{f}^{\prime}\rangle|o_{f}\rangle}
\end{equation}

mediated by $\mathcal{E}$.  Here $|\mathcal{A}^{\mathit{i}}\rangle$ and $|\mathcal{E}^{\mathit{i}}\rangle$ refer to unmeasured initial states of the apparatus and the environment, and in line with tradition $\mathcal{A}$ is represented as monitoring the states of a distinct, microscopic quantum system described by basis vectors $\lbrace |s_{k}\rangle\rbrace$.  Because the environment separates $\mathcal{A}$ from $\mathcal{O}$ and prevents their coherent entanglement, the $\mathcal{O-E}$ interaction occurs slightly later than and at some remove from the $\mathcal{A-E}$ interaction; this is reflected in Eqn. 3 by $|e_{f}^{\prime}\rangle$ indicating a time-propagated descendent of $|e_{f}\rangle$.  

The first of these two processes, represented by Eqn. 2, has the same form as the traditional von Neumann chain, with the sole difference that $\mathcal{E}$ has replaced $\mathcal{O}$ in the role of observer.  If no further physical assumptions are made, the choice of basis vectors $|s_{k}\rangle$ remains arbitrary, and the assumption of strong decoherence has no impact on the description of measurement (e.g. \cite{adler03}).  The concept of einselection \cite{zurek03rev, zurek98rough, zurek93rev} is based on the realization that, in any practical observational context, the specification of $|\mathcal{E}^{\mathit{i}}\rangle$ involves the physically-significant assumption of a particular $\mathcal{A-E}$ interaction, e.g. scattering of ambient photons or immersion in a heat bath.  Specification of the $\mathcal{A-E}$ interaction implicitly specifies the states of $\mathcal{A}$ that can be observed non-destructively, and hence implicitly specifies a basis for $\mathcal{A}$, e.g. position in the case of photon scattering or energy in the case of heat-bath immersion.  These implicit specifications can be made explicit by specifying the einselected basis of $\mathcal{A}$, i.e. by re-writing Eqn. 2 as:

\begin{equation}
|\mathcal{S}^{\mathit{i}}\rangle\mathit{(\sum_{k}\lambda^{\prime}_{k}|a_{k}\rangle)}|\mathcal{E}^{\mathit{i}}\rangle\,\rightarrow \mathit{|s_{f}\rangle|a_{f}\rangle|e_{f}\rangle}
\end{equation}

where $|\mathcal{S}^{\mathit{i}}\rangle$ is the unmeasured and for all practical purposes irrelevant initial state of the microscopic system and the $|a_{k}\rangle$ are eigenvectors of the $\mathcal{A-E}$ interaction Hamiltonian $H_{\mathcal{A-E}}$.  From this ``environment as witness'' perspective, the assumption of strong decoherence \textit{does} affect the description of measurement: it fixes the basis for the joint state $|s_{f}\rangle|a_{f}\rangle|e_{f}\rangle$ as the eigenbasis $\lbrace |a_{k}\rangle\rbrace$ of $H_{\mathcal{A-E}}$, requires $|a_{f}\rangle$ to be an eigenstate in this basis, and allows the final environmental state $|e_{f}\rangle$ to be regarded as an ``encoding'' of $|a_{f}\rangle$ in this basis \cite{zurek04, zurek05}.  

Within the environment as witness formulation, Eqn. 3 is merely an afterthought; the observer plays no role in fixing the basis $\lbrace |a_{k}\rangle\rbrace$ \cite{zurek03rev}.  The time-propagated environmental encoding $|e_{f}^{\prime}\rangle$ imposes the final observer state $|o_{f}\rangle$ on $\mathcal{O}$ by einselection \cite{zurek03rev, zurek09rev}.  The exclusive role of the environment in fixing the einselected basis is systematically obscured by the common practice of using a partial trace over $\mathcal{E}$ to effectively collapse Eqns. 3 and 4 back to a ``decohered'' version of Eqn. 1 in which $\lbrace |a_{k}\rangle\rbrace$ replaces $\lbrace |s_{k}\rangle\rbrace$ as the explicit basis.  As noted by Zurek \cite{zurek03rev}, such a partial trace assumes the Born rule and hence also obscures, by treating the Born rule as a mathematical axiom, the physical origin of the probabilities defined over quantum-mechanical observables.   

The reformulation of Eqn. 2, where the choice of basis is unspecified, to Eqn. 4, where it is specified, requires an explicit basis $\lbrace |a_{k}\rangle\rbrace$ for the macroscopic system $\mathcal{A}$.  It therefore rests on the assumption that the macroscopic boundary between $\mathcal{A}$ and $\mathcal{E}$ is well defined, either by an explicit boundary condition on $H_{\mathcal{A-E}}$ or by an explicit specification of matrix elements $\langle a_{i}|H_{\mathcal{A-E}}\mathit{|e_{j}\rangle}$.  However, neither decoherence theory nor its extension to quantum Darwinism provides a method for calculating either macroscopic boundary conditions or explicit matrix elements using only quantum-mechanical assumptions and the results of non-destructive measurements \cite{fields10a, fields10b}; classical assumptions must be employed to distinguish $\mathcal{A}$ from $\mathcal{E}$ in either case.  While the requirement that a distinction between $\mathcal{A}$ and $\mathcal{E}$ must be made in order to define $H_{\mathcal{A-E}}$ has been broadly noted, the requirement that this distinction be stated in classical terms has not.  Zurek, for example, elevates to the status of ``axiom(o)'' of quantum mechanics the assumption that there are ``systems'' distinguishable by boundaries at which decoherence acts (\cite{zurek03rev}, p. 746; \cite{zurek07grand}, p. 3), but fails to note that in the absence of a purely quantum-mechanical method for calculating macroscopic boundary conditions or specifying explicit matrix elements, an instance of this axiom is required for every macroscopic object.  It is not unreasonable to ask why this is the case, to ask why the description of measurement using decoherence requires as an \textit{axiom} the assumption that my old Geiger counter, for example, exists as a system.  In order to answer this question, we ask a different one: what are the physical assumptions implicit in the afterthought, Eqn. 3, that describes the interaction between the environment and an observer?  Answering this question requires specifying what the interaction called ``observation'' is.

If $\mathcal{E}$ imposes strong decoherence, Eqn. 3 describes effectively classical information transfer: the interaction between $\mathcal{O}$ and $\mathcal{E}$ changes the state of $\mathcal{O}$ but leaves $\mathcal{E}$ effectively unaltered.  The final observer state $|o_{f}\rangle$, in particular, is einselected as an eigenstate of the $\mathcal{O-E}$ interaction and hence is effectively classical.  Provisionally assuming that both the $\mathcal{A-E}$ and $\mathcal{O-E}$ interactions are well defined allows Eqn. 3 to be provisionally accepted as a description of classical information transfer and the scenario in the Introduction to be reformulated as a communication \textit{task} in LOCC terms.  Each of the $\mathcal{A}_{\mathit{i}}$ becomes a transmitter of classical messages through a common channel $\mathcal{E}$ to a receiver $\mathcal{O}$.  The $k^{th}$ message from $\mathcal{A}_{\mathit{i}}$ contains a value $r_{ik}$ of the read-out $\mathcal{R}_{\mathit{i}}$, together with an identifier and possibly other information (e.g. operating temperature, expected battery life, etc.) about the state of $\mathcal{A}_{\mathit{i}}$.  The task of $\mathcal{O}$ is to accumulate multiple messages from each sender and then, at the end of the observation period, to report the accumulated values $r_{ik}$, with each value tagged by the identity of the sender.  For simplicity, the messages are assumed to be composed of discrete classical bits received by $\mathcal{O}$ sequentially and without noise or overlap.  While observing the string of bits coming from $\mathcal{E}$, $\mathcal{O}$ must 1) parse them into syntactically well-formed messages identified by sender, and 2) extract from each message the read-out value $r_{ik}$ to be recorded.  These tasks are distinct and must be performed sequentially, as the read-out values transmitted by distinct apparatus $\mathcal{A}_{\mathit{i}}$ and $\mathcal{A}_{\mathit{j}}$ may be indistinguishable; both $\mathcal{A}_{\mathit{i}}$ and $\mathcal{A}_{\mathit{j}}$ may, for example, transmit the read-out value ``2''.  On the assumption that $\mathcal{E}$ has at least the computational complexity of a classical finite state machine, classical automata theory demonstrates that these two tasks are non-trivial and that they cannot be performed without prior information: $\mathcal{O}$ cannot define, from any finite sample of input from $\mathcal{E}$, either the bit patterns that distinguish well-formed messages from each sender or which bits within each well-formed message encode content to be recorded, even if $\mathcal{O}$ is permitted finite diagnostic inputs to the $\mathcal{A}_{\mathit{i}}$ \cite{moore56}.  The parsers required to identify well-formed messages by sender and extract the read-out values must, therefore, be encoded by any system - any ``minimal observer'' - capable of carrying out the task of observation as described.  A classical Bob cannot, therefore, receive all of the instructions necessary for decoding Alice's transmissions from Alice; classical automata theory requires that Bob must encode the parsers that enable the \textit{recognition} of Alice's transmissions in advance.  This requirement for advance ``knowledge'' results solely from the imposition of LOCC; in the absence of direct Alice-Bob entanglement, i.e. in the presence of decoherence in the communication channel, Bob requires classical information relating states of the communication channel to states of Alice before the communication can begin.

The message-parsing functions that a minimal observer must encode in order to recognize messages encoded in $\mathcal{E}$ are summarized in Fig. 1.  As these functions must be available at the outset of the observations, they must be encoded in executable form within $|\mathcal{O}^{\mathit{r}}\rangle$.  The specification of $|\mathcal{O}^{\mathit{r}}\rangle$ in Eqn. 3, therefore, like the specification of $|\mathcal{E}^{\mathit{i}}\rangle$ in Eqn. 2, involves a significant physical assumption.  By specifying $|\mathcal{O}^{\mathit{r}}\rangle$, one is implicitly assuming that the self-interaction $H_{\mathcal{O-O}}$ of $\mathcal{O}$ is a Hamiltonian oracle, in the sense defined by Farhi and Gutmann \cite{farhi96}, that implements both a set $\lbrace P_{i}\rbrace$ of parsers, each of which recognizes environmental encodings of one of the $\mathcal{A}_{\mathit{i}}$, and a further set $\lbrace R_{ik}\rbrace$ of parsers that recognize and extract the particular values $r_{ik}$ of the $\mathcal{R}_{\mathit{i}}$ from the output of the $P_{i}$.  The sense in which a physical system driven by a Hamiltonian oracle $H_{\mathcal{O-O}}$ ``implements'' these parsers, or implements any algorithm, is the familiar sense of classical computer science; the parsers constitute a ``virtual machine'' running on $\mathcal{O}$ \cite{tan76}.  As Farhi and Gutmann point out, the dynamics of any physical system ``designed to solve a specified problem'' - i.e. any system that can be described as executing an algorithm - can be represented as a Hamiltonian oracle (\cite{farhi96}, p. 2403); hence the notion that virtual machines are implemented by Hamiltonian oracles is completely general.  Specifying a virtual machine does not, however, specify the Hamiltonian oracle that implements it.  Any virtual machine can be implemented by an infinite number of distinct physical systems \cite{tan76}, and hence by an infinite number of distinct Hamiltonian oracles.

\psset{xunit=1cm,yunit=1cm}
\begin{pspicture}(0,0)(16,11)
\put(5.2,10.2){$|\mathcal{O}^{\mathit{r}}\rangle$}
\put(5.5,10){\vector(0,-1){1}}
\pspolygon(4,8)(5.5,9)(7,8)(5.5,7)
\put(4.5,7.9){$\mathcal{A}_{\mathit{1}}$ signal?}
\put(7,8){\line(1,0){1.5}}
\put(7.5,8.2){No}
\put(8.5,8){\line(0,-1){.3}}
\psdot(8.5,7.3)
\psdot(8.5,7)
\psdot(8.5,6.7)
\put(8.5,6.3){\vector(0,-1){.3}}
\pspolygon(7,5)(8.5,6)(10,5)(8.5,4)
\put(7.5,4.9){$\mathcal{A}_{\mathit{N}}$ signal?}
\put(10,5){\vector(1,0){1.5}}
\put(10.5,5.2){No}
\put(5.5,7){\vector(0,-1){0.5}}
\pspolygon(4,5.5)(4,6.5)(7,6.5)(7,5.5)
\put(4.5,5.9){Extract $\mathcal{R}_{\mathit{1}}$}
\put(5.5,5.5){\vector(0,-1){4}}
\put(8.5,4){\vector(0,-1){.5}}
\pspolygon(7,2.5)(7,3.5)(10,3.5)(10,2.5)
\put(7.5,2.9){Extract $\mathcal{R}_{\mathit{N}}$}
\put(8.5,2.5){\vector(0,-1){1}}

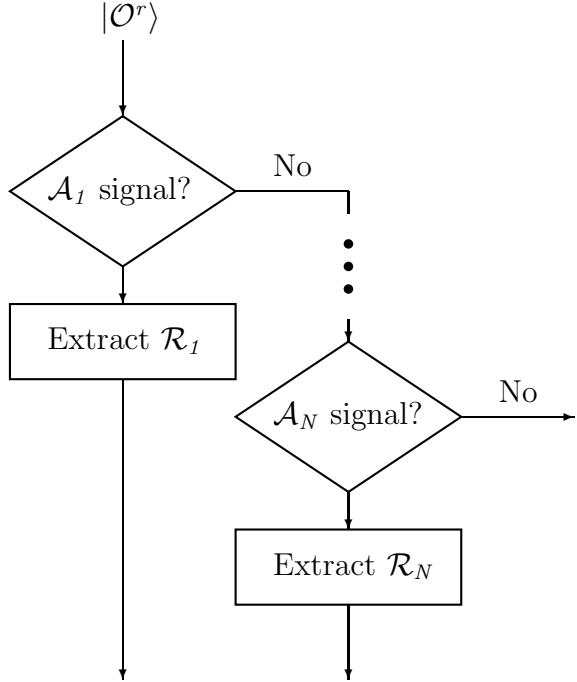
\put(2.5,0.5){\textit{Fig. 1: Functions required to parse a message from $\mathcal{E}$}}
\end{pspicture}

As are Equations 3 and 4, Fig. 1 is consistent with $\mathcal{O}$ making a single, effectively instantaneous observation.  However, $\mathcal{O}$'s task as described is to make multiple observations over an extended period, and then to report the accumulated results.  These task requirements impose additional architectural constraints on the minimal $\mathcal{O}$.  In particular, performing multiple observations requires both that $\mathcal{O}$ maintain prior observations in a reliable memory, and that $\mathcal{O}$ return reliably to $|\mathcal{O}^{\mathit{r}}\rangle$ at the end of each observation cycle.  These requirements are summarized in the functional decomposition of the extended observation task shown in Fig. 2.  Extended observation is a loop from $|\mathcal{O}^{\mathit{r}}\rangle$ back to $|\mathcal{O}^{\mathit{r}}\rangle$ with $N + 1$ conditional branches.  Three functional requirements for executing this loop are evident:

\begin{enumerate}
\item $\mathcal{O}$ must incorporate a reliable memory that stores not only the parsers $\lbrace P_{i}\rbrace$ and $\lbrace R_{ik}\rbrace$ required to recognize inputs and extract their reportable components, but also the results $r_{ik}$, ordered by cycle and tagged by source, of executing these parsers.
\item $\mathcal{O}$ must implement a functional architecture capable of executing loops and input-dependent conditional branching.
\item $\mathcal{O}$'s functional architecture must specifically return $\mathcal{O}$ to $|\mathcal{O}^{\mathit{r}}\rangle$ following each memory-write operation.
\end{enumerate}

For a fixed, finite set of $\mathcal{A}_{\mathit{i}}$ and a fixed number of observation cyles, the architecture of a classical finite-state machine with a fixed memory will meet these requirements.  However, a general observer capable of carrying out the task of Fig. 2 for any observation period and any possible finite set of $\mathcal{A}_{\mathit{i}}$ cannot be assumed to pre-allocate memory for all possible observation periods or to encode parsers for all possible $\mathcal{A}_{\mathit{i}}$; such an observer must, therefore, incorporate a dynamically-allocatable memory and a capability to acquire and store specifications of recognizers for new $\mathcal{A}_{\mathit{i}}$ not previously encountered.  A minimal observer capable of carrying out extended observations of multiple macroscopic systems must, therefore, implement a functional architecture functionally equivalent to a classical Turing Machine \cite{tan76}.  Specifying $|\mathcal{O}^{\mathit{r}}\rangle$ in Eqn. 3 therefore involves the significant physical assumption that the self-interaction $H_{\mathcal{O-O}}$ is a Hamiltonian oracle that implements a Turing-equivalent functional architecture.

\psset{xunit=1cm,yunit=1cm}
\begin{pspicture}(0,0)(16,16)
\put(14.5,15.1){\vector(-1,0){10.2}}
\put(3.2,15){$|\mathcal{O}^{\mathit{r}}\rangle$}
\put(3.5,14.7){\vector(0,-1){0.7}}
\pspolygon(2,13)(3.5,14)(5,13)(3.5,12)
\put(2.5,12.9){observing?}
\put(5,13){\vector(1,0){4.5}}
\put(5.5,13.2){No}
\put(3.5,12){\vector(0,-1){1}}
\pspolygon(2,10)(3.5,11)(5,10)(3.5,9)
\put(2.5,9.9){$\mathcal{A}_{\mathit{1}}$ signal?}
\put(5,10){\line(1,0){1.5}}
\put(5.5,10.2){No}
\put(6.5,10){\line(0,-1){.3}}
\psdot(6.5,9.3)
\psdot(6.5,9)
\psdot(6.5,8.7)
\put(6.5,8.3){\vector(0,-1){.3}}
\pspolygon(5,7)(6.5,8)(8,7)(6.5,6)
\put(5.5,6.9){$\mathcal{A}_{\mathit{N}}$ signal?}
\psdot(14.5,7)
\put(8,7){\vector(1,0){6.5}}
\put(8.5,7.2){No}
\pspolygon(9.5,12.5)(9.5,13.5)(12.5,13.5)(12.5,12.5)
\put(9.7,12.9){Report records}
\put(11,12.5){\vector(0,-1){1.5}}
\pspolygon(9.5,10)(9.5,11)(12.5,11)(12.5,10)
\put(9.8,10.4){Flush records}
\put(11,10){\line(0,-1){.5}}
\psdot(14.5,9.5)
\put(11,9.5){\vector(1,0){3.5}}
\put(3.5,9){\vector(0,-1){0.5}}
\pspolygon(2,7.5)(2,8.5)(5,8.5)(5,7.5)
\put(2.5,7.9){Extract $\mathcal{R}_{\mathit{1}}$}
\put(3.5,7.5){\vector(0,-1){5}}
\pspolygon(2,1.5)(2,2.5)(5,2.5)(5,1.5)
\put(2.5,1.9){Record $\mathcal{R}_{\mathit{1}}$}
\put(3.5,1.5){\line(0,-1){0.2}}
\put(3.5,1.3){\line(1,0){11}}
\put(14.5,1.3){\line(0,1){13.8}}
\psdot(9.5,2.2)
\psdot(9.5,1.9)
\psdot(9.5,1.6)
\put(6.5,6){\vector(0,-1){.5}}
\pspolygon(5,4.5)(5,5.5)(8,5.5)(8,4.5)
\put(5.5,4.9){Extract $\mathcal{R}_{\mathit{N}}$}
\put(6.5,4.5){\vector(0,-1){0.5}}
\pspolygon(5,3)(5,4)(8,4)(8,3)
\put(5.5,3.4){Record $\mathcal{R}_{\mathit{N}}$}
\put(6.5,3){\line(0,-1){0.5}}
\psdot(14.5,2.5)
\put(6.5,2.5){\vector(1,0){8}}

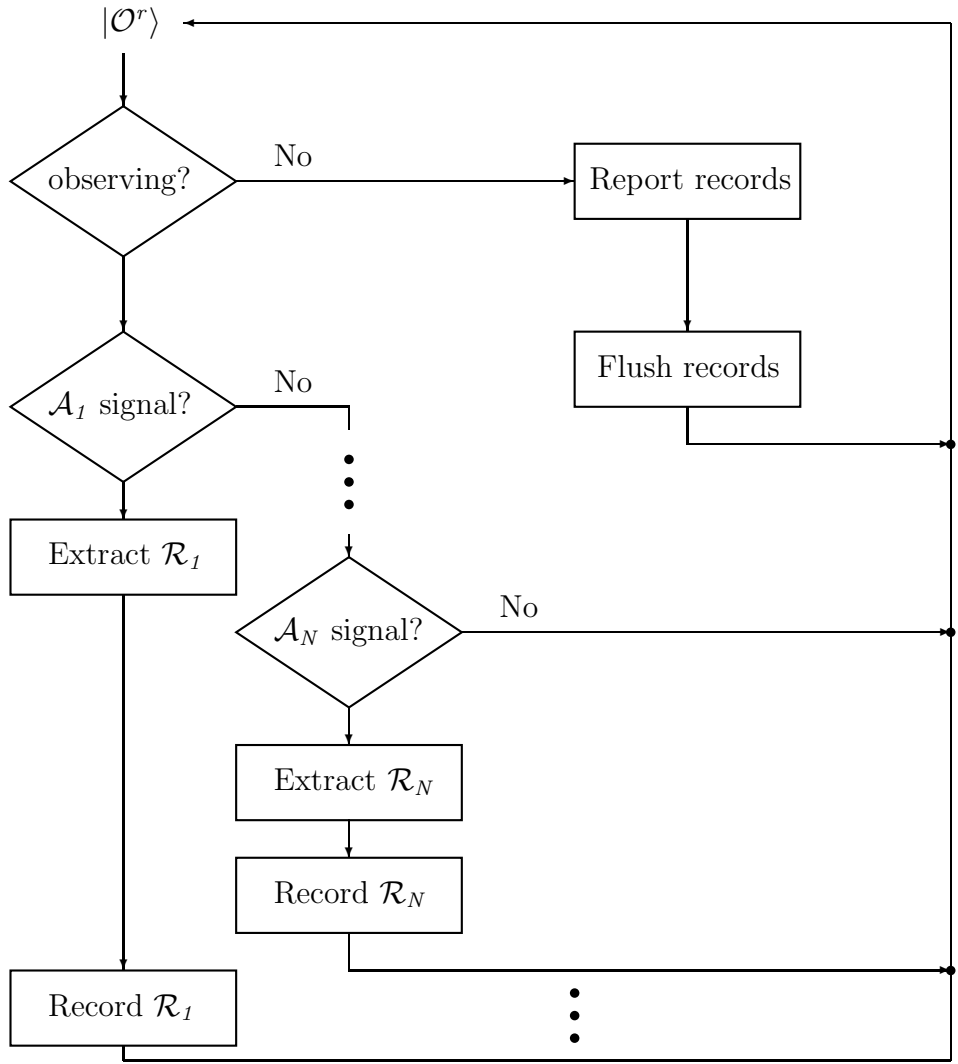
\put(2.5,0.5){\textit{Fig. 2: Functional decomposition of multiple-observation task}}
\end{pspicture}

Turing-equivalent functional architectures permit a functional distinction between the ``processor'' that executes algorithms and the associated ``memory'' that stores both executable specifications of algorithms and the results of running them.  This functional distinction allows the physical assumptions inherent in Eqn. 3 can be made more explicit.  Let $\mathcal{M^{O}}$ refer to the physical memory components of $\mathcal{O}$ that change state on each observation cycle.  Indicating the state of $\mathcal{M^{O}}$ following the $j^{th}$ memory-write operation as $|\mathcal{M^{O}}_{\mathit{j}}\rangle$, Eqn. 3 can be re-written as:

\begin{equation}
|e_{f}^{\prime}\rangle ( |\mathcal{O}^{\mathit{r}}\rangle|\mathcal{M^{O}}_{\mathit{j}}\rangle ) \,\rightarrow \mathit{|e_{f}^{\prime}\rangle} ( |\mathcal{O}^{\mathit{r}}\rangle|\mathcal{M^{O}}_{\mathit{j+1}}\rangle )
\end{equation}

where the arrow represents not einselection by $H_{\mathcal{O-E}}$ as in Eqn. 3, but rather the action over the $j^{th}$ observation cycle of $H_{\mathcal{O-O}}$.  It is clear from Eqn. 5 that the action of $H_{\mathcal{O-O}}$ clones $|\mathcal{O}^{\mathit{r}}\rangle$; therefore by the no-cloning theorem \cite{wooters82}, $|\mathcal{O}^{\mathit{r}}\rangle$ not only encodes classical information but must be a classical state.  It is implicit in Eqn. 5 and clear from Fig. 2 that the action of $H_{\mathcal{O-O}}$ also clones the states of the $j$ previously-written cells of $\mathcal{M^{O}}$; therefore by the no-cloning theorem $\mathcal{M^{O}}$ must be a classical memory.  The classicality results obtained by Tegmark \cite{tegmark00} by analyzing decoherence rates in mammalian nervous systems thus apply equally to all minimal observers, solely in virtue of the task requirements of extended observation.  Writing down ``$|\mathcal{O}^{\mathit{r}}\rangle$'' in a von Neumann chain, therefore, by itself implies the assumption of a \textit{classical} minimal observer.  As shown in the next section, this result depends only on $H_{\mathcal{O-O}}$ being a Hamiltonian oracle implementing message parsers on a Turing-equivalent functional architecture, and is independent of any assumptions about the dynamical structure of the environment beyond an absence of bias for a particular coordinate system, including in particular the assumption of decoherence.

\section{Quantum measurement by minimal observers}

The analysis of observation as a task carried out in the previous section began with the assumptions that $\mathcal{E}$ imposed strong decoherence and that both the $\mathcal{A-E}$ and $\mathcal{O-E}$ interactions were well-defined.  Equation 5 explicitly distinguishes $\mathcal{O}$ from $\mathcal{E}$ and hence requires a well-defined $\mathcal{O-E}$ interaction, i.e. a well-defined $\mathcal{O-E}$ boundary; however, Eqn. 5 does not appeal to einselection and does not mention $\mathcal{A}$.  It is therefore worth asking if the assumption of a well-defined $\mathcal{A-E}$ interaction, and in particular of a decohering $\mathcal{A-E}$ interaction can be relaxed. 

Suppose that ``axiom(o)'' \cite{zurek03rev, zurek07grand} is dropped and no ``systems'' other than $\mathcal{O}$ are assumed.  Decoherence then acts simultaneously at every possible compact (let us assume for simplicity) boundary that does not overlap $\mathcal{O}$, and pointer states of every possible compact assemblage of quantum degrees of freedom, including all single degrees of freedom, not contained within $\mathcal{O}$ are simultaneously encoded in the state of $\mathcal{E}$.  In this case, Equations 2 and 4 become superfluous, the notion that $|e_{f}^{\prime}\rangle$ is an encoding of the states of some particular systems $\mathcal{A}_{\mathit{i}}$ can be dropped, and Eqn. 5 can be re-written:

\begin{equation}
|\mathcal{E}^{\mathit{i}}\rangle ( |\mathcal{O}^{\mathit{r}}\rangle|\mathcal{M^{O}}_{\mathit{j}}\rangle ) \,\rightarrow |\mathcal{E}^{\mathit{i}}\rangle ( |\mathcal{O}^{\mathit{r}}\rangle|\mathcal{M^{O}}_{\mathit{j+1}}\rangle )
\end{equation}

where as in Eqn. 5 the arrow represents the action of $H_{\mathcal{O-O}}$ and therefore does not affect $|\mathcal{E}^{\mathit{i}}\rangle$.  Here $|\mathcal{E}^{\mathit{i}}\rangle$ is interpreted as simultaneously encoding the pointer states of every possible compact assemblage of quantum degrees of freedom outside of $\mathcal{O}$, but the \textit{physical situation} it describes is no different from that described by Eqn. 5.  Re-writing the equation changes nothing about the ambient photon field, for example, and nothing about the dispositions of matter previously, but no longer, singled out and referred to as the ``$\mathcal{A}_{\mathit{i}}$''.  If the physical situation is unchanged, however, the response of $\mathcal{O}$ to that situation must be unchanged as well.  The parsers implemented by $\mathcal{O}$ produce the same outputs as before, and $\mathcal{O}$ reports the same observational results, even though the boundaries of the $\mathcal{A}_{\mathit{i}}$ remain unspecified.  Zurek's ``axiom(o)'' thus adds nothing to the physical situation.  The system boundaries at which decoherence acts do not need to be well-defined within the theory, and assuming well-defined boundaries and hence well-defined $\mathcal{A-E}$ interactions is not required to accept Eqn. 3 or its derivatives Eqns. 5 and 6.  Decoherence, however, is defined as a process that acts at system-environment boundaries \cite{zurek03rev, schloss04, schloss07}.  If well-defined $\mathcal{A-E}$ boundaries are unnecessary to the description of measurement, decoherence and einselection are unnecessary as well.  If the assumptions of well-defined $\mathcal{A-E}$ boundaries and hence well-defined $\mathcal{A-E}$ interactions, decoherence, and einselection - the assumptions that enabled the classical treatment of the observation task in Sect. 2 - are all dropped, a ``systems-free'' formulation of the $\mathcal{O-E}$ interaction results.  The specification of $|\mathcal{O}^{\mathit{r}}\rangle$, with the concommitant assumption of a Turing-equivalent functional architecture implemented by a Hamiltonian oracle, by itself enforces LOCC in this systems-free formulation.

In the systems-free formulation, $\mathcal{O}$'s ability to report the states of external ``systems'' $\mathcal{A}_{\mathit{i}}$ rests entirely on the action of $\mathcal{O}$'s parsers as implemented by the Hamiltonian oracle $H_{\mathcal{O-O}}$.  Let $O_{ik} = H_{\mathcal{O-E}} \circ P_{i} \circ R_{ik}$, where ``$\circ$'' denotes composition and ``$P_{i}$'' and ``$R_{ik}$'' are used to denote the projections of $H_{\mathcal{O-O}}$ that constitute the Hamiltonian oracles implementing these parsers.  Two facts about the set of operators $\lbrace O_{ik}\rbrace$ are evident from Fig. 2.  First, on any given cycle of observation, exactly one of the $O_{ik}$ will return a value $r_{ik}$, or else none will.  The $O_{ik}$ are, therefore, effectively orthogonal and together resolve the identity over $\mathcal{E}$.  Second, the values $r_{ik}$ can be represented as finite strings of classical bits, and hence, without loss of generality, as positive integers.  The $\lbrace O_{ik}\rbrace$ are therefore a POVM defined over $\mathcal{E}$.  For each $i$, moreover, the subset $\lbrace O_{ik}\rbrace$ is a POVM that detects all and only the states of $\mathcal{E}$ that are reported by $\mathcal{O}$ to encode signals from the ``system'' $\mathcal{A}_{\mathit{i}}$ implicitly defined by the parser $P_{i}$; hence the subset $\lbrace O_{ik}\rbrace$ is an \textit{observable} for $\mathcal{A}_{\mathit{i}}$.  Unlike traditional quantum-mechanical observables such as Position, the $O_{ik}$ are not observer-independent and do not act on ``systems'' external to the observer.  Instead they act locally on $\mathcal{O}$'s environment.  One of the motivating insights of quantum Darwinism - that observers interact with the local environment, not with distant systems \cite{zurek06, zurek09rev} - thus follows as a consequence of the assumption of minimal observers.

Suppose observers $\mathcal{O}^{\mathit{(1)}}$ and $\mathcal{O}^{\mathit{(2)}}$ implement POVMs $\lbrace O^{(1)}_{ik}\rbrace$ and $\lbrace O^{(2)}_{ik}\rbrace$ respectively, and that:

\begin{equation}
\forall i,k \, \mathrm{and} \, \forall |\mathcal{E}^{\mathit{i}}\rangle, \mathit{O^{(1)}_{ik}}|\mathcal{E}^{\mathit{i}}\rangle = \mathit{O^{(2)}_{ik}}|\mathcal{E}^{\mathit{i}}\rangle = \mathit{r_{ik}}.
\end{equation}

In this case $\lbrace O^{(1)}_{ik}\rbrace$ and $\lbrace O^{(2)}_{ik}\rbrace$ are output-equivalent and $\mathcal{O}^{\mathit{(1)}}$ and $\mathcal{O}^{\mathit{(2)}}$ can be said to share a POVM $\lbrace O_{ik}\rbrace$.  Multiple minimal observers will agree about an $\mathcal{A}_{\mathit{i}}$ present in their shared environment if and only if they share a POVM $\lbrace O_{ik}\rbrace$ that outputs a single set $\lbrace r_{ik}\rbrace$ of values for the observable states of the read-out $\mathcal{R}_{\mathit{i}}$.  The notion of ``objectivity'' has been defined within quantum Darwinism as:

\begin{quotation}
``A property of a physical system is \textit{objective} when it is:
\begin{list}{\leftmargin=2em}
\item
1. simultaneously accessible to many observers,
\item
2. who are able to find out what it is without prior knowledge about the system of interest, and 
\item
3. who can arrive at a consensus about it without prior agreement."
\end{list}
\begin{flushright}
(p. 1 of \cite{zurek04}; p. 3 of \cite{zurek05})
\end{flushright}
\end{quotation}

A set of minimal observers that share POVMs will agree, using this definition, that the states of the $\mathcal{A}_{\mathit{i}}$ they jointly observe are objective and hence effectively classical.

It was assumed at the outset that the environment $\mathcal{E}$ contains more degrees of freedom than any observer $\mathcal{O}$; hence $\mathcal{O}$ cannot, in principle, implement a recognizer for any environmental state $|\mathcal{E}^{\mathit{i}}\rangle$.  Any minimal observer is, therefore, objectively ignorant as a consequence of architecture of the state of the environment and cannot, even in principle, determine whether environmental states encountered at two distinct times are the same.  A minimal observer cannot, moreover, infer anything about the dynamics of $\mathcal{E}$ except that they are sufficient to generate the values of the $r_{ik}$ obtained with the observables available to $\mathcal{O}$.  Any finite sequence of observed values $r_{ik}$ can be generated by an infinite number of distinct finite state machines \cite{moore56}; any one of which could be implemented by an infinite number of distinct Hamiltonian oracles.  The standard quantum-mechanical representation of any total environmental state $|\mathcal{E}^{\mathit{i}}\rangle$ as an arbitrary superposition in an arbitrary basis can thus be viewed as a consequence of the objective ignorance of minimal observers \cite{fields11}.

As shown by Zurek \cite{zurek05env}, provable ignorance on the part of observers generates the Born rule.  Zurek's proof of the Born rule from envariance requires a controversial (e.g. \cite{schlossfine07}) \textit{a priori} assumption of a course-graining defined over the system of interest on which a classical probability measure sums to unity.  For a minimal observer as defined here, an orthogonal course-graining over any observable system $\mathcal{A}_{\mathit{i}}$ is provided by the POVM $\lbrace O_{ik}\rbrace$, which as a resolution of the identity also provides a classical probability measure guaranteed by the architecture of $\mathcal{O}$ to sum to unity.  With this observer-specific course-graining and the notation:

\begin{equation}
|\mathcal{R}_{\mathit{i}}\rangle = \mathit{\sum_{k} \xi_{k} |r_{ik}\rangle}
\end{equation}

where $r_{ik}$ is the value of $|\mathcal{R}_{\mathit{i}}\rangle$ recognized and extracted by $\mathcal{O}$'s $k^{th}$ $\mathcal{A}_{\mathit{i}}$-specific POVM component $O_{ik}$, the Born rule becomes:

\textit{Theorem} (\textbf{Born Rule}): $P_{k} = |\xi_{k}|^{2}$.

The proof depends both on $\mathcal{A}_{\mathit{i}}$ and $\mathcal{O}$ interacting with $\mathcal{E}$ in an unbiased way and on $\mathcal{O}$'s inability to distinguish environmental states $|e_{j}^{k}\rangle$ that encode $|r_{ik}\rangle$, and follows that given in \cite{zurek05env}.

\textit{Proof}:
Let $C$ be an ancillary ``counter'' of the environmental states $|e_{j}^{k}\rangle$ that encode each $|r_{ik}\rangle$.  Assuming an $\mathcal{A_{\mathit{i}}-E}$ interaction that is unbiased with respect to coordinate system and hence with respect to $k$, for a suitably large $C$ the fraction $M_{k}$ of states in $C$ that encode each $|r_{ik}\rangle$ can be chosen as $M_{k} = \vert\xi_{k}\vert^{2}$ to an arbitrarily good approximation.  Suppose that $\mathcal{O}$ sequentially observes every state in $C$.  At any stage in this process, the probability that the next state observed will result in a report of $r_{ik}$ is $\frac{M_{k}}{\sum_{k} M_{k}} = M_{k} = \vert\xi_{k}\vert^{2}$ as required.  $\square$

\section{Conclusion}

This paper has examined the physical assumptions implicit in describing a temporally-extended situation as ``observation'', picking out a particular system $\mathcal{O}$ as an ``observer'', and specifying an initial state $|\mathcal{O}^{\mathit{r}}\rangle$ as the observer ``ready state''.  It has focused on a characterization of observation as a \textit{task} described in classical terms.  It shows that if this classical characterization of observation as a temporally-extended task is accepted, classical automata theory defines a minimal observer $\mathcal{O}$: the self-interaction $H_{\mathcal{O-O}}$ of $\mathcal{O}$ must be a Hamiltonian oracle that implements a Turing-equivalent functional architecture, and the ready state $|\mathcal{O}^{\mathit{r}}\rangle$ must encode executable specifications of parsers that recognize and extract reportable classical values from signals transmitted from the environment by the $\mathcal{O-E}$ interaction.  Minimal observers defined in this way are very familiar: any laboratory data-collection system running on a general-purpose computer is a minimal observer in this sense.

The paper then examines the physical assumptions that must be made, and the physical assumptions that can be dropped, in the course of describing the physical interactions between a macroscopic quantum system $\mathcal{O}$ that satisfies the definition of a minimal observer and the surrounding environment $\mathcal{E}$.  It shows that the assumption of a minimal observer by itself enforces LOCC.  It shows that the parsers implemented by a minimal observer fully determine both the external quantum systems $\mathcal{A}_{\mathit{i}}$ with which $\mathcal{O}$ can report interactions and the read-out values $r_{ik}$ that can be reported as observational results.  It shows that these parsers composed with the Hamiltonian $H_{O-E}$, which must be well-defined but which need not be specified explicitly, constitute POVMs and hence observables specific to the external systems that $\mathcal{O}$ is capable of observing.  It shows that multiple minimal observers will agree about observational outcomes and hence about the ``objectivity'' of external systems if and only if they share POVMs for those systems.  Finally it shows, under the assumption that physical interactions are not biased toward particular coordinate systems and hence basis vectors, that the observations reported by $\mathcal{O}$ will comply with the Born Rule.  None of these demonstrations require any assumptions beyond the absence of bias to particular coordinate systems about the interactions between $\mathcal{E}$ and any external systems, or any assumptions about einselection or decoherence.  Hence these latter assumptions do not need to be made: observation can be described in systems-free terms.

Minimal observers as defined here are quantum systems that exhibit intrinsically classical behavior.  All observed physics, however, is quantum; as Fuchs \cite{fuchs02} somewhat poetically puts it, the world has ``zing'', and it is this ``zing'' that distinguishes the predictions of quantum mechanics from, in Fuchs' purely subjective framework, those of classical Bayesian inference.  In the present treatment, the source of this ``zing'' is the obligate, architecturally-enforced ignorance that generates the Born rule.  Minimal observers do not and cannot implement recognizers for the environmental states $|\mathcal{E}^{\mathit{i}}\rangle$ with which they interact; hence even if $|\mathcal{E}^{\mathit{i}}\rangle$ were not a quantum state and could be cloned, $\mathcal{O}$ could not, even in principle, recognize it as cloned.  Minimal observers cannot, therefore, replicate initial states of $\mathcal{E}$, even if these states are classical; the probabilities given by the Born rule provide the full extent of the knowledge $\mathcal{O}$ can obtain about physical causation.  It is acknowledgement of this obligate ignorance of $|\mathcal{E}^{\mathit{i}}\rangle$ on the part of observers, not any fact concerning dynamics, that distinguishes quantum from classical mechanics in the systems-free formulation \cite{fields11}. 

Treating the observer as a quantum system that implements a classical functional architecture represents a profound departure from the von Neumann - Everett tradition of treating the observer as an essentially unstructured quantum system, the states of which are entirely determined by its entanglement with the environment.  The difference between the present treatment and the von Neumann - Everett tradition is interpretational, i.e. semantic: the tradition treats observation as a physical interaction to be specified quantum mechanically, while the present treatment defines observation as a physical interaction that completes a \textit{task} that is specified classically.  The measurement problem arises in the traditional treatment because the description of observation as a physical interaction conflicts with intuitions regarding what counts as an ``outcome'', a ``report'' or an ``experience'' of observation.  These latter terms are not quantum-mechanical terms, but rather classical terms that describe observation as a task to be performed.  Hence the traditional treatment demands a ``Heisenberg cut'' between quantum and classical descriptions, or else an infinite tree of ``branches'' that specify classical outcomes.  The present treatment resolves the question of the ``boundary'' between quantum and classical descriptions in the simple and familiar way that computer science resolves the question of the ``boundary'' between a computer and its user interface: the task of observation is performed by a classical virtual machine running on a quantum physical system.  The observational capabilities of this virtual machine, like the task it performs, are described classically.  The underlying physical dynamics are described quantum mechanically, as the composition of the Hamiltonian oracle $H_{\mathcal{O-O}}$ with the observer-environment interaction $H_{\mathcal{O-E}}$.  In this treatment, there is no measurement problem; "outcomes" and "reports" are virtual-machine states that are related to the underlying quantum dynamics by the Born rule.

The Hamiltonian oracles that implement minimal observers share a fundamental requirement: they must implement reliable memory.  The traditional treatment of observation is not concerned with the memory of the observer; indeed the traditional treatment can do without observers.  In the current treatment, reliable memory is a fundamental assumption; without reliable memory, the classical task of observation cannot be performed.  Physical interactions, however, cannot clone quantum states; hence \textit{perfectly} reliable memory is impossible.  Observation and hence science is feasible to the extent that observers can, as Zurek \cite{zurek03rev} puts it, ``readily consult the content of their memory'' and regard the results of such consultation as sufficiently reliable to proceed.

\section*{Acknowledgement}

The comments of an anonymous referee contributed significantly to the clarity of this paper.

\end{document}